\documentclass{l4dc2027}
\usepackage{times}
\usepackage{booktabs,multirow,array,microtype,mathtools}
\usepackage{graphicx,xcolor,array}
\usepackage{enumitem}
\usepackage{tikz}
\usetikzlibrary{fit,backgrounds}
\title[Adaptive Koopman Predictive Control without Persistent Excitation]{Adaptive Koopman Predictive Control without Persistent Excitation\\ via Matrix Normal Invariant Priors}
\author{%
 \Name{Mircea Lazar} \Email{m.lazar@tue.nl}\\
 \addr Department of Electrical Engineering, Eindhoven University of Technology, The Netherlands}
\newcommand{\R}{\mathbb{R}}
\newcommand{\Ko}{\mathcal K}
\newcommand{\Kn}{\mathcal K_0}
\newcommand{\Tr}{\operatorname{tr}}
\newcommand{\col}{\operatorname{col}}
\newcommand{\norm}[1]{\left\lVert #1\right\rVert}
\newcommand{\T}{\mathsf T}
\newcommand{\ph}{\varphi}
\newcommand{\Qnorm}[1]{\lVert #1\rVert_{Q^{-1}}}
\newcommand{\doi}[1]{doi:\,\href{https://doi.org/#1}{\mbox{\nolinkurl{#1}}}}
\makeatletter
\renewcommand\ps@jmlrtps{\let\@mkboth\@gobbletwo\def\@oddhead{}\def\@evenhead{}\def\@oddfoot{}\def\@evenfoot{}}
\makeatother
\begin{document}
\maketitle
\begin{abstract}
Finite-dimensional Koopman models are increasingly used in predictive control, but their accuracy remains limited by truncation errors and offline identification data. Adaptive Koopman predictive control promises to address this limitation, but typically requires closed-loop persistent excitation (PE). Since this is difficult to achieve in practice, in this paper we develop an adaptive generalized Koopman (GeKo) predictive controller that does not require PE. To this end, we formulate a recursive matrix regression in fixed lifted coordinates that extends the Kalman filter with back-to-the-prior forgetting to matrix-valued operators and uses a matrix normal invariant prior, centered at the operator identified offline. For any regressor sequence, and thus under any feedback law, we prove that the covariance remains between two multiples of the prior covariance and that the estimation error is input-to-state stable.  On an academic example, a differential-drive robot and a power system with plant changes and little excitation, the invariant-prior estimator yields the lowest overall tracking error among the compared estimators and prevents covariance windup.
\end{abstract}
\section{Introduction}
Model predictive control (MPC) needs a model that remains accurate over the state and input closed-loop trajectories. Koopman predictors identified offline are attractive for this purpose \citep{Williams2015,Korda2018,Strasser2026}, but they lose accuracy when a system parameter, such as the load for a robot or the surface under a vehicle, changes. Updating the predictor online is the natural remedy. It is also delicate, since a well-performing controller produces data with little excitation, a classical difficulty in adaptive control that can cause parameter drift and bursting \citep{Anderson1985,AnnaswamyFradkov2021,Annaswamy2023}. Furthermore, classical forgetting then inflates the covariance in the unexcited directions.

Adaptive Koopman predictive control methods that address this typically target \emph{linear} Koopman models \citep{Korda2018} of the form $z^+=Az+Bu$, with $z=\Psi_x(x)$ the lifted state. Recursive updates, which process each new sample at a fixed cost, were developed for extended dynamic mode decomposition \citep{Sinha2019} and for Koopman MPC \citep{Calderon2021}. Online dynamic mode decomposition discounts older samples to track time-varying systems \citep{Zhang2019}. Other schemes re-identify the operator from a sliding window of recent data \citep{DelRio2025}. Since closed-loop data are poorly exciting, constant-trace normalization, dead-zone gating, and subspace forgetting \citep{Lai2024} were compared for the covariance of recursive linear Koopman model learning \citep{Gu2026}. Concurrent learning for lifted linear models \citep{Najarzadeh2026} combines a gradient update with a memory of informative past regressors, following \citep{Chowdhary2010,Chowdhary2014}, so that stored data compensate for missing excitation. Other approaches learn corrections to a nominal model over a sliding window \citep{Singh2024,Sah2026}, update a neural lifting online \citep{Uchida2024}, or add steady-state constraints to the learned predictor \citep{Cen2026}. The available guarantees are conditional on the online data. Recursive least squares requires persistent excitation for uniform exponential convergence \citep{Ljung1983}, and covariance regulation ensures only that the covariance remains bounded \citep{Gu2026}. Concurrent learning relaxes persistent excitation to finite excitation, in the form of a rank or contraction condition on the stored regressors \citep{Chowdhary2010,Sah2026}. Both conditions are difficult to enforce in closed loop. Furthermore, linear  Koopman models are subject to an error floor even for infinitely many state observables \citep{Lazar2025GeKo}.

The generalized Koopman (GeKo) operator for control \citep{Lazar2025GeKo} is defined on tensor products of state and input observables, and its finite-dimensional approximations are linear in the operator matrix and converge. Hence, standard recursive estimators can be employed. To eliminate the stringent requirement of persistent excitation in closed loop, we adopt a \emph{back-to-the-prior forgetting} approach \citep{VanVaerenbergh2012}. Reverting the estimate toward a prior value is the leakage modification of classical adaptive control, which keeps the estimation error bounded without excitation \citep{IoannouSun1996,Landau2011}, at the price of an offset \citep{Bitmead1990}. Recursions that also revert the covariance are known from estimation and tracking \citep{Rosenberg1973,KulhavyZarrop1993}. In particular, back-to-the-prior forgetting was developed for kernel recursive least squares \citep{VanVaerenbergh2012}; therein, its analysis for linear regression was left open. The same recursion was recently used with a forgetting rate that is learned online, for linear predictors on neural features \citep{Titsias2024} and for autoregressive with exogenous input (ARX) predictors in data-driven predictive control \citep{Chiuso2026}, where the prior, which is not forgotten over time, was aptly termed an \emph{invariant prior}.

In this paper we extend back-to-the-prior forgetting to matrix-valued Koopman operators by means of a \emph{matrix normal invariant prior}, with the offline GeKo operator as mean and a fixed forgetting rate, which makes it possible to derive deterministic bounds on the estimation error. The contributions of this paper are as follows: \emph{(i)}~we formulate GeKo adaptation as one matrix regression with a matrix normal invariant prior centered at the offline operator, for which back-to-the-prior forgetting admits a single covariance for all rows of the operator, of the dimension of the regressor rather than of the number of operator entries; \emph{(ii)}~we derive explicit bounds for the resulting \emph{invariant-prior estimator} (IPE) that require no excitation, including input-to-state stability (ISS) of the estimation error; \emph{(iii)}~we compare recursive least-squares-type and invariant-prior estimators in closed loop on three challenging benchmarks.

\section{Preliminaries}\label{sec:prelim}
Consider a time-varying system $x_{k+1}=f_k(x_k,u_k)$ with $x_k\in\R^{n_x}$ and $u_k\in\R^{n_u}$. Select state and input dictionaries $\Psi_x:\R^{n_x}\to\R^{n_z}$ and $\Psi_u:\R^{n_u}\to\R^{n_v}$, whose first entries are equal to one and which stay fixed during adaptation, and write $z=\Psi_x(x)$. Then the first entry of the regressor $\ph=z\otimes\Psi_u(u)\in\R^{p}$, with $p=n_zn_v$ and $\otimes$ the Kronecker product, is one, and the constant observable is predicted exactly by $[1\ 0\ \cdots\ 0]$. Hence, the  GeKo predictor \citep{Lazar2025GeKo} becomes
\begin{equation}\label{eq:geko}
  z^+=\col(1,\Ko\,\ph),\qquad \ph=z\otimes\Psi_u(u),
\end{equation}
where $\Ko\in\R^{(n_z-1)\times p}$ collects the remaining rows, which are estimated.

The GeKo predictor \eqref{eq:geko} is nonlinear in $(x,u)$, but linear in $\Ko$. Bilinear predictors correspond to $\Psi_u(u)=\col(1,u)$, that is, to GeKo with a finite input subspace, and the linear Koopman predictor is recovered by the same form with the regressor $\ph=\col(z,u)$. Thus the analysis applies to linear and bilinear Koopman models, with the distinction that for fixed dictionaries they differ in the regression residual. Given $N_0$ offline transitions $(x_i,u_i,x_i^+)$, $i=1,\ldots,N_0$, let the columns of $\Phi_0$ and $Y_0$ be the regressors $\Psi_x(x_i)\otimes\Psi_u(u_i)$ and the lifted successor states $\Psi_x(x_i^+)$ without their constant entry, respectively. Then the nominal operator is obtained via the Khatri--Rao  regression \citep{Lazar2025GeKo}:
\begin{equation}\label{eq:K0}
 \Kn=Y_0\Phi_0^\T(\Phi_0\Phi_0^\T+\varepsilon I)^{-1},\qquad \varepsilon>0 .
\end{equation}

\subsection{Recursive least squares estimators in Kalman filter form}\label{sec:estimators}
At each time $k\ge1$, the measured state $x_k$, together with $x_{k-1}$ and $u_{k-1}$, provides the regressor $\ph_k=\Psi_x(x_{k-1})\otimes\Psi_u(u_{k-1})\in\R^p$ and the measurement $y_k\in\R^{n_z-1}$, which contains the entries of $\Psi_x(x_k)$ except the constant one. By \eqref{eq:geko}, these data satisfy $y_k\approx\Ko\ph_k$, which is linear in $\Ko$. Hence, recursive least squares (RLS) can be applied to all rows of $\Ko$. As in the Kalman filter, we write the measurement model
\begin{equation}\label{eq:meas_model}
 y_k=\Ko_k\ph_k+v_k,\qquad \sigma_k^2=\sigma^2\bigl(1+\norm{\ph_k}^2\bigr),
\end{equation}
where $\Ko_k$ is the operator at time $k$, $v_k$ has zero mean and covariance $\sigma_k^2U$ given the data up to time $k-1$, with $U\succ0$, and $\sigma^2>0$ is a noise level. The factor $1+\norm{\ph_k}^2$ is equivalent to dividing the regressor and the measurement by $\sqrt{1+\norm{\ph_k}^2}$, which is the normalization of adaptive laws \citep[Sec.~4.3.5]{IoannouSun1996}. As in \citep{AndersonMoore1979}, $\Ko_{k|k-1}$ and $P_{k|k-1}$ denote the estimate and its covariance given the measurements up to time $k-1$, and $\Ko_{k|k}$ and $P_{k|k}$ denote them after the measurement $y_k$. All rows of $\Ko$ share one covariance $P\succ0$, since they have the same regressor $\ph_k$ and the noise covariance $\sigma_k^2U$ has the same scale $\sigma_k^2$ for all entries of $y_k$. We formulate the measurement update as the regularized least squares problem
\begin{equation}\label{eq:ls}
 \Ko_{k|k}=\arg\min_{\Ko}\ J_k(\Ko),\quad J_k(\Ko)=\Tr\bigl((\Ko-\Ko_{k|k-1})P_{k|k-1}^{-1}(\Ko-\Ko_{k|k-1})^\T\bigr)+\norm{y_k-\Ko\ph_k}^2/\sigma_k^2,
\end{equation}
see, e.g., \citep[Eq.~(2.4.1)]{Kailath2000} for a similar formulation. Setting the gradient of $J_k$ to zero yields the information form
\begin{equation}\label{eq:info}
 \Ko_{k|k}P_{k|k}^{-1}=\Ko_{k|k-1}P_{k|k-1}^{-1}+y_k\ph_k^\T/\sigma_k^2,\qquad
 P_{k|k}^{-1}=P_{k|k-1}^{-1}+\ph_k\ph_k^\T/\sigma_k^2,
\end{equation}
and the Sherman--Morrison formula \citep[Eq.~(3.8.2)]{Meyer2000} gives
\begin{equation}\label{eq:meas}
\begin{aligned}
 L_k&=\frac{P_{k|k-1}\ph_k}{\sigma_k^2+\ph_k^\T P_{k|k-1}\ph_k},\\
 \Ko_{k|k}&=\Ko_{k|k-1}+\bigl(y_k-\Ko_{k|k-1}\ph_k\bigr)L_k^\T,\\
 P_{k|k}&=P_{k|k-1}-\frac{P_{k|k-1}\ph_k\ph_k^\T P_{k|k-1}}{\sigma_k^2+\ph_k^\T P_{k|k-1}\ph_k},
\end{aligned}
\end{equation}
which is the measurement update of the Kalman filter \citep[Lemma~9.3.2]{Kailath2000}, with the gain $L_k\in\R^p$ shared by all rows of $\Ko$. In the analysis, we use the innovation $e_k=y_k-\Ko_{k|k-1}\ph_k$ and $S_k=\sigma_k^2+\ph_k^\T P_{k|k-1}\ph_k$, so that $L_k=P_{k|k-1}\ph_k/S_k$. All estimators in this paper use \eqref{eq:meas} and differ only in the time update from $(\Ko_{k|k},P_{k|k})$ to $(\Ko_{k+1|k},P_{k+1|k})$, which implements the forgetting. Table~\ref{tab:estimators} lists RLS with exponential forgetting \citep{Ljung1983} and its variants used for linear Koopman predictors in \citep{Gu2026}. They use $\sigma^2=1$, $\Ko_{1|0}=\Kn$, $P_{1|0}=rI$ with $r>0$ and the forgetting factor $\lambda=\exp(-T/\tau)$, where $T$ is the sampling period and $\tau$ is a time constant.

\begin{table}[t]\centering\footnotesize
\caption{Forgetting in RLS-type estimators, combined with the measurement update \eqref{eq:meas}, with DZ-CS \citep{Gu2026} and SIFt \citep{Lai2024}.}
\label{tab:estimators}
\setlength{\tabcolsep}{5pt}\renewcommand{\arraystretch}{1.15}
\begin{tabular}{@{}lll@{}}\toprule
Estimator & Forgetting & Covariance without excitation \\\midrule
RLS & $\Ko_{k+1|k}=\Ko_{k|k}$,\quad $P_{k+1|k}=P_{k|k}/\lambda$ & unbounded (windup) \\
DZ-CS & RLS, after $P_{k|k}\leftarrow P_{k-1|k-1}$ if $\norm{e_k}\le\delta\sigma_k$ & unbounded while $\norm{e_k}>\delta\sigma_k$ \\
SIFt & $\Ko_{k+1|k}=\Ko_{k|k}$,\quad $P_{k+1|k}=P_{k|k}+\frac{1-\lambda}{\lambda\,\ph_{k+1}^\T P_{k|k}^{-1}\ph_{k+1}}\ph_{k+1}\ph_{k+1}^\T$ & bounded above and below\\
\bottomrule\end{tabular}
\end{table}

\section{The invariant-prior estimator and adaptive GeKo predictive control}\label{sec:main}
The estimator considered in this paper combines the measurement update \eqref{eq:meas} with the back-to-the-prior forgetting introduced in \citep[Eq.~(16)]{VanVaerenbergh2012} for a scalar output and a Gaussian posterior with a zero-mean prior. We extend the forgetting to the matrix-valued operator $\Ko$, with a single prior covariance $Q$ for all rows and the offline operator $\Kn$ as prior mean, as follows.

\begin{definition}[Matrix normal invariant prior] \citep[Sec.~2]{GuptaNagar2000}\label{def:prior}
A random matrix $X\in\R^{n\times p}$ has a matrix normal distribution with mean $M$, row covariance $U\succ0$ and column covariance $\Sigma\succ0$, denoted by $X\sim\mathcal{MN}(M,U,\Sigma)$, if $\operatorname{vec}(X)$, which stacks the columns of $X$, is Gaussian with mean $\operatorname{vec}(M)$ and covariance $\Sigma\otimes U$. Given $U\succ0$ and $Q\succ0$, the prior $\mathcal{MN}(\Kn,U,Q)$ is invariant if $\Ko_k\sim\mathcal{MN}(\Kn,U,Q)$ implies $\Ko_{k+1}\sim\mathcal{MN}(\Kn,U,Q)$.
\end{definition}
The entries of $X$ then satisfy $\operatorname{Cov}(X_{ij},X_{kl})=U_{ik}\Sigma_{jl}$, and $X\sim\mathcal{MN}(M,U,\Sigma)$ means $X^\T\sim N_{p,n}(M^\T,\Sigma\otimes U)$ in the notation of \citep[Def.~2.2.1, Thm.~2.3.1]{GuptaNagar2000}. Under this prior, each operator row is Gaussian with covariance proportional to $Q$, and $U$ describes correlation among rows. Given a persistence factor $\rho=\exp(-T/\tau)\in(0,1)$, the \emph{invariant-prior estimator} (IPE) for the matrix-valued operator combines the measurement update \eqref{eq:meas} with the time update
\begin{equation}\label{eq:ipe_time}
 \Ko_{k+1|k}=\Kn+\rho(\Ko_{k|k}-\Kn),\qquad P_{k+1|k}=\rho^2P_{k|k}+(1-\rho^2)Q,\qquad k\ge1,
\end{equation}
where $\Ko_{1|0}=\Kn$ and $P_{1|0}=Q$. The IPE corresponds to the measurement model \eqref{eq:meas_model} together with the operator model
\begin{equation}\label{eq:opmodel}
 \Ko_{k+1}=\Kn+\rho(\Ko_k-\Kn)+\sqrt{1-\rho^2}\,W_k,
\end{equation}
where the unknown operator $\Ko_k$ is modeled as a random matrix, $W_k\sim\mathcal{MN}(0,U,Q)$ and $v_k=\sigma_k\varepsilon_k$ with $\varepsilon_k\sim\mathcal N(0,U)$. The sequences $\{W_k\}_{k\ge1}$ and $\{\varepsilon_k\}_{k\ge1}$ are each independent over time, and $\Ko_1$, $\{W_k\}_{k\ge1}$ and $\{\varepsilon_k\}_{k\ge1}$ are mutually independent. Since $\ph_k$ and $\sigma_k$ depend only on the data up to time $k-1$, $v_k$ is Gaussian with covariance $\sigma_k^2U$ given these data.
\begin{proposition}[Invariant prior]\label{prop:prior}
\emph{(i)}~The prior $\mathcal{MN}(\Kn,U,Q)$ is invariant for \eqref{eq:opmodel}. \emph{(ii)}~Let $\Ko_1\sim\mathcal{MN}(\Kn,U,Q)$ and let each regressor $\ph_k$ depend only on the data available at time $k-1$. Then, for all $k\ge1$ and given the measurements up to time $k$, it holds that $\Ko_k\sim\mathcal{MN}(\Ko_{k|k},U,P_{k|k})$ and $\Ko_{k+1}\sim\mathcal{MN}(\Ko_{k+1|k},U,P_{k+1|k})$, where $\Ko_{k|k}$, $P_{k|k}$ are given by \eqref{eq:meas} and $\Ko_{k+1|k}$, $P_{k+1|k}$ by \eqref{eq:ipe_time}.
\end{proposition}
\begin{proof}
\emph{(i)}~Let $\Ko_k\sim\mathcal{MN}(\Kn,U,Q)$. Then $\rho(\Ko_k-\Kn)\sim\mathcal{MN}(0,U,\rho^2Q)$ and $\sqrt{1-\rho^2}\,W_k\sim\mathcal{MN}(0,U,(1-\rho^2)Q)$ are independent, so that by \eqref{eq:opmodel} their sum satisfies $\Ko_{k+1}-\Kn\sim\mathcal{MN}(0,U,Q)$. \emph{(ii)}~We show by induction that, given the data up to time $k-1$, $\Ko_k\sim\mathcal{MN}(\Ko_{k|k-1},U,P_{k|k-1})$. For $k=1$, this is the assumption on $\Ko_1$, with $\Ko_{1|0}=\Kn$ and $P_{1|0}=Q$. Assume it holds for some $k$. Since $\ph_k$ is known and $v_k$ is independent of $\Ko_k$, the measurement $y_k=\Ko_k\ph_k+v_k$ is jointly Gaussian with $\Ko_k$, with mean $\Ko_{k|k-1}\ph_k$ and covariance $(\ph_k^\T P_{k|k-1}\ph_k+\sigma_k^2)U=S_kU$. Conditioning on $y_k$, the factor $U$ cancels in the gain $L_k$, and $\Ko_k\sim\mathcal{MN}(\Ko_{k|k},U,P_{k|k})$, where $\Ko_{k|k}$ and $P_{k|k}$ are given by \eqref{eq:meas}, see \citep[Thm.~16.4]{WestHarrison1997}, conditional on $\Sigma=U$ and with $\Theta_t=(\Ko_k-\Kn)^\T$. Then, since $W_k$ is independent of $\Ko_k$ and of the data up to time $k$, \eqref{eq:opmodel} and \eqref{eq:ipe_time} give $\Ko_{k+1}-\Ko_{k+1|k}=\rho(\Ko_k-\Ko_{k|k})+\sqrt{1-\rho^2}\,W_k\sim\mathcal{MN}(0,U,\rho^2P_{k|k}+(1-\rho^2)Q)$, that is, $\Ko_{k+1}\sim\mathcal{MN}(\Ko_{k+1|k},U,P_{k+1|k})$, which completes the induction.
\end{proof}
By Proposition~\ref{prop:prior}, the recursion \eqref{eq:meas}, \eqref{eq:ipe_time} propagates the conditional mean and covariance of $\Ko_k$, that is, it is the Kalman filter for \eqref{eq:meas_model}, \eqref{eq:opmodel}. Since $U$ does not appear in \eqref{eq:meas}, \eqref{eq:ipe_time}, the estimator can run without knowing the correlation $U$.

The model \eqref{eq:meas_model}, \eqref{eq:opmodel} is a working model that motivates the recursion, since noise passed through nonlinear dictionaries is in general neither Gaussian nor independent of the regressor. By \eqref{eq:ipe_time}, forgetting moves the estimate toward the offline operator $\Kn$ and its covariance toward $Q$. Moreover, the matrix-valued operator is estimated with a single $p\times p$ covariance $P$, whose update requires $O(p^2)$ operations per sample, whereas a Kalman filter on $\operatorname{vec}(\Ko)$ propagates an $(n_z-1)p\times(n_z-1)p$ covariance, with $(n_z-1)^2p^2$ entries to update per sample.

Two prior covariances are considered. IPE-iso uses $Q=q_0I$, for which only the ratio $r=q_0/\sigma^2$ affects the estimate. IPE-data uses $Q=q_0\,\Omega\operatorname{diag}(\alpha_1,\ldots,\alpha_p)\Omega^\T$, where $\Omega$ contains the eigenvectors of the normalized offline information matrix $G_0=N_0^{-1}\sum_{i=1}^{N_0}\ph_i\ph_i^\T/(1+\norm{\ph_i}^2)$, with $\ph_i$ the columns of $\Phi_0$, and the weights $\alpha_i>0$ average to one and decrease with the eigenvalues $\lambda_i$ of $G_0$, so that directions that are poorly excited offline receive a larger prior variance. Weights proportional to $(\lambda_i+r_0)^{-1}$ give the shape of the regularized offline posterior covariance, where $r_0>0$ limits the condition number of $Q$, which enters the constant $c$ of Theorem~\ref{thm:iss}.

The results derived next hold for any regressor sequence, and thus for any feedback law, and they do not require the working model to hold, so that $P$ serves only as a weighting and gain matrix. Let $\Ko_k^\star$ be an arbitrary comparator sequence, such as the best operator for the plant at time $k$, and define the estimation error, the residual, the prior offset and the variation
\begin{equation}\label{eq:comparator}
E_k=\Ko_k^\star-\Ko_{k|k-1},\qquad \xi_k=y_k-\Ko_k^\star\ph_k,\qquad B_k=\Ko_k^\star-\Kn,\qquad D_{k+1}=\Ko_{k+1}^\star-\Ko_k^\star .
\end{equation}
The prior offset $B_k$ measures how far the plant is from the offline model, the variation $D_{k+1}$ how fast it changes, and the residual $\xi_k$ collects the measurement noise, including the state noise passed through the dictionaries, and the structural model error. By \eqref{eq:ipe_time}, $E_{k+1}=\rho(\Ko_k^\star-\Ko_{k|k})+(1-\rho)B_k+D_{k+1}$, and by \eqref{eq:meas}, $\Ko_k^\star-\Ko_{k|k}=E_k-e_kL_k^\T$ with $e_k=E_k\ph_k+\xi_k$. Hence, the estimation error obeys
\begin{equation}\label{eq:errsys}
E_{k+1}=E_kA_k^\T-\rho\,\xi_kL_k^\T+(1-\rho)B_k+D_{k+1},\qquad A_k=\rho\bigl(I-L_k\ph_k^\T\bigr),\qquad k\ge1,
\end{equation}
which is a linear time-varying system with the matrix state $E_k\in\R^{(n_z-1)\times p}$ and the inputs $B_k$, $D_{k+1}$ and $\xi_k$, whose time variation enters only through $\ph_k$ and $P_{k|k-1}$. Each row of $E_k$ is the error of one row of the operator, and all rows evolve with the same matrix $A_k$. For $W\succ0$, let $\norm{M}_W=\Tr(MWM^\T)^{1/2}$ denote the $W$-weighted Frobenius norm. For the state of \eqref{eq:errsys} define
\begin{equation}\label{eq:eta}
V_k:=\norm{E_k}_{P_{k|k-1}^{-1}}=\Tr\bigl(E_kP_{k|k-1}^{-1}E_k^\T\bigr)^{1/2},
\end{equation}
as a candidate ISS Lyapunov function.

\begin{lemma}[Covariance bounds]\label{lem:cov}
For all $k\ge1$, it holds that $(1-\rho^2)Q\preceq P_{k|k-1}\preceq Q$ and $0\prec P_{k|k}\preceq P_{k|k-1}$, $P_{k|k}\preceq P_{k+1|k}$.
\end{lemma}
\begin{proof}
If $P_{k|k-1}\succ0$, then \eqref{eq:info} gives $P_{k|k}\succ0$ and \eqref{eq:meas} gives $P_{k|k}\preceq P_{k|k-1}$. Hence, if $(1-\rho^2)Q\preceq P_{k|k-1}\preceq Q$, which holds for $k=1$ since $P_{1|0}=Q$, then \eqref{eq:ipe_time} gives $(1-\rho^2)Q\preceq P_{k+1|k}\preceq\rho^2Q+(1-\rho^2)Q=Q$ and $P_{k+1|k}-P_{k|k}=(1-\rho^2)(Q-P_{k|k})\succeq0$. The claim follows by induction.
\end{proof}
The upper bound in Lemma~\ref{lem:cov} prevents covariance windup, and the lower bound prevents the covariance from collapsing, so that the estimator remains able to adapt. Both bounds are deterministic and hold for any regressor sequence, since \eqref{eq:ipe_time} is a convex combination of $P_{k|k}$ and $Q$. The next lemma derives an instrumental bound on $V_{k+1}$.

\begin{lemma}\label{lem:energy}
For all $k\ge1$, it holds that
\begin{equation}\label{eq:onestep}
V_{k+1}\le\rho\sqrt{V_k^2-\frac{\norm{e_k}^2}{S_k}+\frac{\norm{\xi_k}^2}{\sigma_k^2}}+c_B\Qnorm{B_k}+\frac{\Qnorm{D_{k+1}}}{\sqrt{1-\rho^2}},\qquad c_B=\sqrt{\frac{1-\rho}{1+\rho}}.
\end{equation}
\end{lemma}
\begin{proof}
We first consider the measurement update. By \eqref{eq:info}, the quadratic cost $J_k$ in \eqref{eq:ls} has the stationary point $\Ko_{k|k}$ and the quadratic term $\Tr(\Ko P_{k|k}^{-1}\Ko^\T)$. Hence, completing the squares gives $J_k(\Ko)=J_k(\Ko_{k|k})+\Tr\bigl((\Ko-\Ko_{k|k})P_{k|k}^{-1}(\Ko-\Ko_{k|k})^\T\bigr)$. By \eqref{eq:meas}, $\Ko_{k|k}-\Ko_{k|k-1}=e_kL_k^\T$ and $y_k-\Ko_{k|k}\ph_k=(\sigma_k^2/S_k)e_k$, with $P_{k|k-1}^{-1}L_k=\ph_k/S_k$, so that $J_k(\Ko_{k|k})=\norm{e_k}^2/S_k$, which is the minimum of the cost \citep[Eq.~(2.4.8)]{Kailath2000}. Since $J_k(\Ko_k^\star)=V_k^2+\norm{\xi_k}^2/\sigma_k^2$ by \eqref{eq:comparator}, \eqref{eq:eta}, evaluating at $\Ko=\Ko_k^\star$ gives
\begin{equation}\label{eq:energy}
\Tr\bigl((\Ko_k^\star-\Ko_{k|k})P_{k|k}^{-1}(\Ko_k^\star-\Ko_{k|k})^\T\bigr)=V_k^2-\norm{e_k}^2/S_k+\norm{\xi_k}^2/\sigma_k^2 .
\end{equation}
Next, we consider the time update, which gives $E_{k+1}=\rho(\Ko_k^\star-\Ko_{k|k})+(1-\rho)B_k+D_{k+1}$. The map $M\mapsto\norm{M}_{P_{k+1|k}^{-1}}$ is a norm \citep[Sec.~5.2]{Meyer2000}, which equals $V_{k+1}$ at $M=E_{k+1}$. Since inversion reverses the order of positive definite matrices \citep[Ch.~7]{HornJohnson2013}, Lemma~\ref{lem:cov} gives $P_{k+1|k}^{-1}\preceq P_{k|k}^{-1}$ and $P_{k+1|k}^{-1}\preceq Q^{-1}/(1-\rho^2)$. Then the triangle inequality and \eqref{eq:energy} yield \eqref{eq:onestep}, where $(1-\rho)/\sqrt{1-\rho^2}=c_B$.
\end{proof}

\begin{definition}[Exponential ISS]\label{def:iss}
The system \eqref{eq:errsys} is exponentially input-to-state stable (ISS), uniformly in the regressor sequence, if there exist $c\ge1$, $\nu\in[0,1)$ and $\gamma_B,\gamma_D,\gamma_\xi\ge0$ such that, for any regressor sequence and any comparator, and for all $k>k_0\ge1$,
\begin{equation}\label{eq:issdef}
\norm{E_k}_F\le c\,\nu^{\,k-k_0}\norm{E_{k_0}}_F+\gamma_B\max_{k_0\le j<k}\Qnorm{B_j}+\gamma_D\max_{k_0\le j<k}\Qnorm{D_{j+1}}+\gamma_\xi\max_{k_0\le j<k}\frac{\norm{\xi_j}}{\sigma_j}.
\end{equation}
\end{definition}
With $x_k=\operatorname{vec}(E_k^\T)$, which stacks the rows of $E_k$, the system \eqref{eq:errsys} becomes $x_{k+1}=(I\otimes A_k)x_k-\rho\,(I\otimes L_k)\xi_k+(1-\rho)\operatorname{vec}(B_k^\T)+\operatorname{vec}(D_{k+1}^\T)$, and $\norm{x_k}=\norm{E_k^\T}_F=\norm{E_k}_F$, since the Frobenius norm is invariant under transposition. Hence, Definition~\ref{def:iss} is the standard ISS of this vector system \citep{JiangWang2001}, with an exponential class-$\mathcal{KL}$ function and linear gains, uniformly in the initial time and in the regressor sequence. Moreover, since the maxima in \eqref{eq:issdef} are taken over the past window, the bound holds along every realization of the noise, it has no offset, and $E_k$ converges to zero if the inputs converge to zero.

\begin{theorem}[ISS of the estimation error]\label{thm:iss}
The system \eqref{eq:errsys} is exponentially ISS, uniformly in the regressor sequence, with $\nu=\rho$, $c=\bigl(\norm{Q}/((1-\rho^2)\lambda_{\min}(Q))\bigr)^{1/2}$, $\gamma_B=\bigl(\norm{Q}/(1-\rho^2)\bigr)^{1/2}$, $\gamma_D=\gamma_B/(1-\rho)$ and $\gamma_\xi=\rho\,\gamma_B$.
\end{theorem}
\begin{proof}
Let $k>k_0\ge1$, fix the regressor sequence, and thus $P_{j|j-1}$, $L_j$ and $A_j$, and let $\beta$, $d$ and $\kappa$ denote the maxima of $\Qnorm{B_j}$, $\Qnorm{D_{j+1}}$ and $\norm{\xi_j}/\sigma_j$ over $k_0\le j<k$. By Lemma~\ref{lem:cov}, $\norm{Q}^{-1}I\preceq Q^{-1}\preceq P_{j|j-1}^{-1}\preceq Q^{-1}/(1-\rho^2)$, which gives
\begin{equation}\label{eq:normeq}
\norm{E_j}_F\le\norm{Q}^{1/2}V_j,\qquad V_j\le\norm{E_j}_F\big/\sqrt{(1-\rho^2)\lambda_{\min}(Q)},\qquad j\ge1.
\end{equation}
Define $\bar E_j$ and $\tilde E_j$ for $j\ge k_0$ by
\begin{equation*}
\bar E_{j+1}=\bar E_jA_j^\T+(1-\rho)B_j+D_{j+1},\quad \bar E_{k_0}=E_{k_0},\qquad
\tilde E_{j+1}=\tilde E_jA_j^\T-\rho\,\xi_jL_j^\T,\quad \tilde E_{k_0}=0 .
\end{equation*}
Adding the two recursions gives \eqref{eq:errsys} with the same initial condition, so that $E_j=\bar E_j+\tilde E_j$. Since \eqref{eq:energy} is an identity in $E_k$ and $\xi_k$, the proof of Lemma~\ref{lem:energy} applies to both recursions, with $\xi=0$ for $\bar E$ and $B=D=0$ for $\tilde E$. With $\bar V_j$ and $\tilde V_j$ defined as in \eqref{eq:eta}, and $\norm{e_j}^2/S_j\ge0$, \eqref{eq:onestep} gives for $k_0\le j<k$
\begin{equation}\label{eq:isslyap}
\bar V_{j+1}\le\rho\bar V_j+c_B\,\beta+\frac{d}{\sqrt{1-\rho^2}},\qquad \tilde V_{j+1}^2\le\rho^2\bigl(\tilde V_j^2+\kappa^2\bigr).
\end{equation}
Iterating \eqref{eq:isslyap} with $\sum_{i\ge0}\rho^i=1/(1-\rho)$, $\sum_{i\ge0}\rho^{2i}=1/(1-\rho^2)$ and $c_B/(1-\rho)=1/\sqrt{1-\rho^2}$, and using $V_k\le\bar V_k+\tilde V_k$ by the norm triangle inequality, gives
\begin{equation}\label{eq:iss}
V_k\le\rho^{\,k-k_0}V_{k_0}+\frac{1}{\sqrt{1-\rho^2}}\Bigl(\beta+\frac{d}{1-\rho}+\rho\,\kappa\Bigr).
\end{equation}
Bounding $\norm{E_k}_F$ by $V_k$ and $V_{k_0}$ by $\norm{E_{k_0}}_F$ via \eqref{eq:normeq} yields \eqref{eq:issdef} with the stated constants. Moreover, \eqref{eq:onestep} and $\sqrt{x^2+y^2}\le x+y$ for $x,y\ge0$ give $V_{j+1}\le\rho V_j+\rho\norm{\xi_j}/\sigma_j+c_B\Qnorm{B_j}+\Qnorm{D_{j+1}}/\sqrt{1-\rho^2}$, so that, with \eqref{eq:normeq}, $V_j$ is an exponential ISS Lyapunov function for \eqref{eq:errsys}.
\end{proof}

By Theorem~\ref{thm:iss}, the estimation error is robust to the prior offset, the plant variation and the residual, uniformly in the regressor sequence, and thus without persistent excitation. Note that Theorem~\ref{thm:iss} asserts boundedness, rather than identification without excitation. Such boundedness is qualitatively known for leakage \citep[Thm.~8.5.4]{IoannouSun1996}, whereas here the gains are explicit in $\rho$. Moreover, the bound \eqref{eq:issdef} is deterministic and it holds along every trajectory, whereas the Kalman filter interpretation of Proposition~\ref{prop:prior} provides statistical properties that require the working model to hold. To the best of our knowledge, Theorem~\ref{thm:iss} is the first exponential ISS result with explicit gains for back-to-the-prior forgetting. The gain $1/\sqrt{1-\rho^2}$ of the prior offset in \eqref{eq:iss} is a worst-case bound on the price of the prior, which acts as a bias if the plant remains far from $\Kn$ \citep[Remark~8.4.1]{IoannouSun1996}. If the plant returns to the offline model from time $k_0$ on, then $B_j=D_{j+1}=0$ and \eqref{eq:iss} gives $V_k\le\rho^{\,k-k_0}V_{k_0}+\rho\kappa/\sqrt{1-\rho^2}$, so that the estimate returns to a neighborhood of $\Kn$ whose size is set by the residual. Theorem~\ref{thm:iss} bounds the estimation error, not the closed-loop trajectories, since $\xi_k$ may depend on the state.

RLS does not have this property. If the comparator is constant, the regressors remain in a subspace $\mathcal M$ after time $k_0$ and $P_{k_0|k_0}\zeta\perp\mathcal M$, then $L_j^\T\zeta=0$ and $P_{j|j}\zeta=\lambda^{-(j-k_0)}P_{k_0|k_0}\zeta$ for all $j>k_0$. Hence, the error in the direction $\zeta$ is not reduced while the covariance grows without bound \citep[Sec.~4.3.6]{IoannouSun1996}, which explains estimator windup \citep{AstromWittenmark1995}.
\begin{remark}[Relation with previous works]\label{rem:rel}
For $\Kn=0$, a scalar output and without the normalization in \eqref{eq:meas_model}, the IPE reduces to the estimator of \citep{VanVaerenbergh2012} with a linear kernel, and to the Kalman filter of \citep{Chiuso2026}. For $\Kn=0$, $Q=q_0I$, $U=I$ and without the normalization, it reduces to the Kalman filter of \citep{Titsias2024}, where independent outputs also share one covariance. In \citep{Titsias2024,Chiuso2026}, the persistence factor $\rho$ is learned online from the predictive likelihood. The IPE differs by the offline operator $\Kn$ as prior mean, by the matrix normal prior with general $Q$ and $U$, and by using a fixed $\rho$, which enables the covariance bounds of Lemma~\ref{lem:cov} and the exponential ISS property of Theorem~\ref{thm:iss} for any regressor sequence.
\end{remark}
\begin{remark}[Model class implications]\label{rem:model}
The model class enters the derived bounds only through the residual $\xi_k$, which consists of the measurement noise and the structural model error. For GeKo predictors, the structural model error can be reduced by richer state and input dictionaries, since the finite-dimensional approximations converge \citep{Lazar2025GeKo}, whereas linear Koopman predictors retain an error floor for any number of state observables.
\end{remark}
\begin{remark}[Further applicability]\label{rem:scope}
The IPE, its $O(p^2)$ update and the derived bounds only use that the predictor is linear in the operator matrix and that all rows share the regressor $\ph_k$. Hence, these properties hold for any predictor $y_k=\Theta\ph_k$ with a prior estimate $\Theta_0$, such as multivariable ARX predictors \citep{Chiuso2026}, linearly parameterized models in adaptive control \citep{IoannouSun1996} and output layers of neural networks adapted online on fixed features \citep{Titsias2024}. In direct adaptive LQR design \citep{Zhao2025}, where $\ph_k=\col(u_k,x_k)$, $P_{k|k}^{-1}$ takes the role of the sample covariance of the data and Lemma~\ref{lem:cov} gives $P_{k|k}^{-1}\succeq Q^{-1}$ without persistently exciting inputs.
\end{remark}

\subsection{Adaptive GeKo predictive control}\label{sec:implementation}
At each time $k$, the adaptive GeKo predictive controller performs the following steps: \emph{(i)}~it measures $x_k$ and forms $y_k$ and $\ph_k=\Psi_x(x_{k-1})\otimes\Psi_u(u_{k-1})$, \emph{(ii)}~computes $(\Ko_{k|k},P_{k|k})$ by \eqref{eq:meas} and $(\Ko_{k+1|k},P_{k+1|k})$ by \eqref{eq:ipe_time}, starting from $\Ko_{1|0}=\Kn$, \emph{(iii)}~forms $\widehat\Ko_k\in\R^{n_z\times p}$ from the known row $[1\ 0\ \cdots\ 0]$ and $\Ko_{k+1|k}$ and solves
\begin{equation}\label{eq:mpc}
\begin{aligned}
\min_{U_k\in\,\mathcal U^H}\ &\sum_{j=1}^{H}w_j\norm{C z_{j|k}-y^r_{j|k}}_{Q_y}^2+\sum_{j=0}^{H-1}\left(\norm{u_{j|k}-u^{r}_{j|k}}_{R}^2+\norm{u_{j|k}-u_{j-1|k}}_{R_\Delta}^2\right)\\
\text{s.t.}\ \ &z_{j+1|k}=\widehat\Ko_k\bigl(z_{j|k}\otimes\Psi_u(u_{j|k})\bigr),\qquad z_{0|k}=\Psi_x(x_k),
\end{aligned}
\end{equation}
and \emph{(iv)}~applies the first input $u_k=u_{0|k}$ of the computed solution. Here, $U_k=\col(u_{0|k},\ldots,u_{H-1|k})$, $H$ is the prediction horizon, $Cz$ are the controlled outputs, which are the states or selected observables, $\mathcal U$ is a box of admissible inputs, $y^r$ and $u^r$ are output and input references, $u_{-1|k}=u_{k-1}$, $w_j=1$ for $j<H$ and $w_H\ge1$ is a terminal weight. For the other estimators of Table~\ref{tab:estimators}, only the time update in \emph{(ii)} changes. For a linear Koopman predictor, $\ph_k=\col(\Psi_x(x_{k-1}),u_{k-1})$ and \eqref{eq:mpc} is a quadratic program (QP). For GeKo, \eqref{eq:mpc} is nonconvex due to the products of lifted states and inputs, and we solve it by a tailored sequential quadratic programming (SQP) method, which implements a Gauss--Newton Hessian and condensing \citep[Sec.~3.1 and~4.3]{Diehl2009}. The tailored SQP method gains numerical efficiency by exploiting the Kronecker structure of the predictor and by using analytical partial derivatives.
The QPs are solved by DAQP \citep{Arnstr2022}, the step is accepted by a backtracking line search on the cost of \eqref{eq:mpc}, and the iteration is initialized with the shifted previous solution.

\section{Numerical evaluation}\label{sec:numerics}\label{sec:benchmarks}
We compare the estimators of Table~\ref{tab:estimators} and the IPE variants on three systems with a nominal phase, a plant change and a return, under sensor noise and little excitation, where the Static predictor keeps $\Kn$, identified from data of the nominal plant. Each run is defined by a random seed, which sets the offline data and the sensor noise.

\begin{figure}[t]\centering
\includegraphics[width=\textwidth]{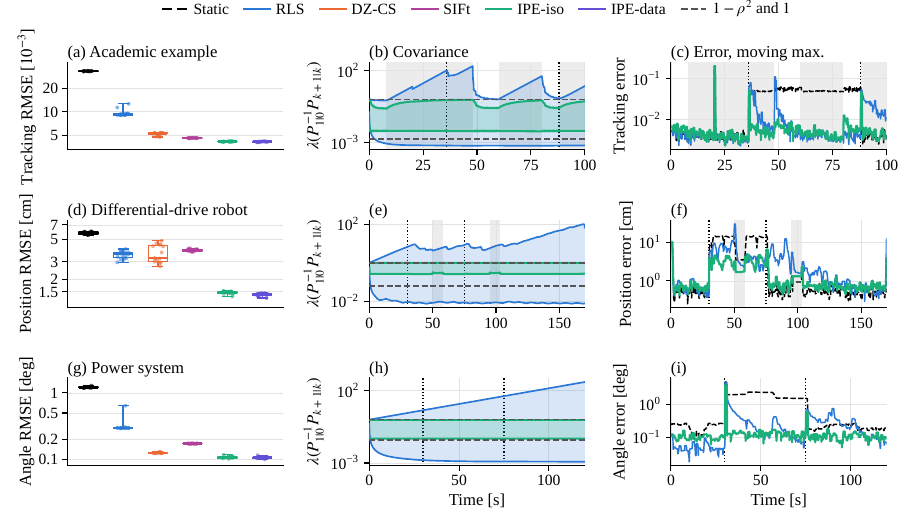}
\caption{GeKo predictor. (a,\,d,\,g) Tracking RMSE over the $15$ runs, with the median as thick line. (b,\,e,\,h) Extreme eigenvalues of $P_{1|0}^{-1}P_{k+1|k}$ and the bounds of Lemma~\ref{lem:cov} for the IPE. (c,\,f,\,i) Maximum absolute tracking error over a sliding window of $0.5$\,s, $1$\,s and $1$\,s, respectively. Vertical dotted lines mark the plant changes, and the reference is constant or the robot stops in the shaded intervals.}\label{fig:systems}
\end{figure}

To compare each estimator at its best, $\tau$, $r$ and $\delta$ were selected per system and predictor by a closed-loop grid search over $5$ seeds that minimizes $\frac15\sum_{s=1}^5(\log\mathrm{RMSE}_s+f_s)$, where $f_s=1$ if run $s$ leaves a safe region and $f_s=0$ otherwise, and all results use $15$ other seeds. This requires simulations of the plant change and only serves the comparison. In practice, the IPE parameters follow from the expected size and duration of the deviation of the operator from $\Kn$, or from the one-step predictive likelihood on offline data of nominal and perturbed plants. The IPE uses $\sigma^2=0.02$ and $q_0=r\sigma^2$, and IPE-data uses $\alpha_i=\tfrac12+\tfrac12\tilde\alpha_i$ with $\tilde\alpha_i=p(\lambda_i+r_0)^{-1}/\sum_{j=1}^p(\lambda_j+r_0)^{-1}$ and $r_0=0.05\Tr(G_0)/p$. Table~\ref{tab:results} reports the median closed-loop tracking errors. Plots (b,\,c,\,e,\,f,\,h,\,i) of Figure~\ref{fig:systems} show the same run for all systems, which was not selected by the IPE performance (it is among the worst of the $15$ runs for the IPE, and for the power system all runs are almost equal).

\begin{table}[t]\centering\footnotesize\vspace{-4pt}
\caption{GeKo predictor: median closed-loop tracking RMSE over $15$ runs, overall and per nominal phase (nom.), changed plant (chg.) and after return (ret.); lowest value per column in bold.}\label{tab:results}
\setlength{\tabcolsep}{3.2pt}
\begin{tabular}{@{}l@{\hspace{6pt}}rrrr@{\hspace{10pt}}rrrr@{\hspace{10pt}}rrrr@{}}\toprule
 & \multicolumn{4}{c}{Academic example, $\times10^{-3}$\hspace{6pt}} & \multicolumn{4}{c}{Robot, position [cm]\hspace{6pt}} & \multicolumn{4}{c}{Power system, angle [deg]} \\
Estimator & overall & nom. & chg. & ret. & overall & nom. & chg. & ret. & overall & nom. & chg. & ret. \\\midrule
Static & $33.5$ & $\mathbf{5.1}$ & $46.0$ & $\mathbf{1.8}$ & $5.86$ & $\mathbf{0.80}$ & $11.32$ & $\mathbf{0.58}$ & $1.220$ & $0.204$ & $1.975$ & $0.198$ \\
RLS & $9.2$ & $\mathbf{5.1}$ & $8.9$ & $16.7$ & $3.59$ & $0.97$ & $5.41$ & $2.84$ & $0.297$ & $\mathbf{0.030}$ & $0.424$ & $0.234$ \\
DZ-CS & $5.3$ & $7.5$ & $3.2$ & $4.4$ & $3.27$ & $1.24$ & $5.74$ & $2.01$ & $0.126$ & $0.038$ & $0.185$ & $0.084$ \\
SIFt & $4.6$ & $5.3$ & $4.0$ & $4.7$ & $3.93$ & $0.94$ & $5.74$ & $3.40$ & $0.173$ & $\mathbf{0.030}$ & $0.250$ & $0.128$ \\
IPE-iso & $\mathbf{4.1}$ & $5.5$ & $\mathbf{3.0}$ & $3.4$ & $1.48$ & $0.84$ & $2.59$ & $0.70$ & $0.109$ & $0.052$ & $0.164$ & $0.051$ \\
IPE-data & $\mathbf{4.1}$ & $5.4$ & $\mathbf{3.0}$ & $3.5$ & $\mathbf{1.41}$ & $0.82$ & $\mathbf{2.47}$ & $0.70$ & $\mathbf{0.107}$ & $0.050$ & $\mathbf{0.162}$ & $\mathbf{0.049}$ \\
\bottomrule\end{tabular}
\end{table}

\emph{Academic example:} We extend the bursting example of \citep[Sec.~3.1.5]{Annaswamy2023}, which goes back to \citep{Anderson1985}, to the bilinear plant $x^+=ax+(b+cx)u$ with $T=20$\,ms. For $z=\col(1,x)$ and $\Psi_u(u)=\col(1,u)$, the regressor is $\ph=\col(1,u,x,xu)$ and the operator row $[0\ \,b\ \,a\ \,c]$ is exact, so that $\xi_k$ is only due to the measurement noise, with standard deviation $0.002$. $\Kn$ is identified from $300$ samples of the nominal plant $(a,b,c)=(0.90,0.15,0.10)$, and the plant equals $(0.80,0.08,0.25)$ from $36$ to $88$\,s. As in \citep{Annaswamy2023}, the input is computed by certainty equivalence. With $\Ko_{k+1|k}=[\hat k_1\ \,\hat k_2\ \,\hat k_3\ \,\hat k_4]$, the predicted state $\hat k_1+\hat k_2u+\hat k_3x_k+\hat k_4x_ku$ is set equal to the reference $x^r_{k+1}$, which gives $u_k=(x^r_{k+1}-\hat k_1-\hat k_3x_k)/(\hat k_2+\hat k_4x_k)$, with $\hat k_2+\hat k_4x_k$ set to $\pm0.02$, with its sign, if it is smaller in magnitude, and saturated to $|u_k|\le2$. The reference is $x^r_k=0.5+0.25\sin(2\pi kT/0.8)+0.12\sin(2\pi kT/0.33)$ for $kT$ in $[0,8)$, $[48,60)$ and $[80,88)$\,s, and $x^r_k=0.5$ otherwise, and the state is increased by $0.2$ at $20$\,s. The challenge is that both plant changes occur while the reference is constant, which causes estimator windup and bursting with exponential forgetting \citep{Anderson1985,AstromWittenmark1995}. The IPE attains the lowest error in all runs, over the whole run and with the changed plant. After the return, its error is the lowest among the adaptive estimators and above that of Static, as the estimate returns to $\Kn$ at the rate $\rho$. The IPE covariance remains in the band of Lemma~\ref{lem:cov}, and with the true operator the estimation error remains below the bound of Theorem~\ref{thm:iss}. In contrast, the RLS covariance winds up, and RLS over-reacts when the excitation resumes at $48$\,s, see Figure~\ref{fig:systems}(b,\,c).

\emph{Differential-drive robot:} The robot model used in \citep{Gu2026}, with $T=50$\,ms, tracks a figure-eight for $170$\,s, with Pacejka slip and Stribeck friction from $30$ to $75$\,s and stops from $50$ to $58$\,s and from $95$ to $103$\,s. We use $z=\col(1,p_x,p_y,\cos\theta,\sin\theta)$ in relative positions, $\Psi_u(u)=\col(1,v,\sin(T\omega)/T,2(1-\cos(T\omega))/T^2)$, which is exact for the nominal forward Euler kinematics, $H=20$, and sensor noise of $4$\,mm and $2$\,mrad. The challenge is to adapt, to keep the adapted model through a stop and to return. Both IPE variants attain a lower error than all other adaptive estimators in all runs and phases. After the return they are close to Static, since the estimate returns to $\Kn$, as predicted after Theorem~\ref{thm:iss}. In contrast, the RLS covariance winds up and the RLS-type estimators do not return to the nominal model, see Figure~\ref{fig:systems}(e,\,f).

\emph{Power system:} We consider the generating station of \citep[Examples~12.2 and~13.1]{Kundur1994}, which is connected to an infinite bus by a transformer and two circuits, with the classical model $\dot\delta=120\pi\,\omega$ and $7\dot\omega=P_m-P_{\max}\sin\delta$ in per unit, where $P_{\max}=1.351$. The valve opening $u\in[0,1]$ sets $P_m=1.2(1-e^{-2u})/(1-e^{-2})$, and \eqref{eq:mpc}, with $T=20$\,ms and $H=30$, tracks a reference for the rotor angle $\delta$ from noisy measurements of $\delta$ and of the speed deviation $\omega$. We use $z=\col(1,\delta,w,\sin\delta,\cos\delta,w\sin\delta,w\cos\delta)$ with $w=100\,\omega$ and $\Psi_u(u)=\col(1,u,\tanh(3u))$. As in \citep[Example~13.1]{Kundur1994}, a three-phase fault on one circuit at $30$\,s, during which $P_{\max}=0$, is cleared after $60$\,ms by isolating the circuit, which gives $P_{\max}=1.102$ until the circuit recloses at $75$\,s. The challenge is a large disturbance, followed by a change that is observed only through the speed. Both IPE variants attain the lowest error in all runs, over the whole run, with the circuit out and after the reclosing, and the lowest peak error after the fault. Before the fault, the RLS-type estimators are more accurate, since the IPE is pulled toward $\Kn$, and the RLS covariance winds up also here, see Figure~\ref{fig:systems}(h,\,i).

\emph{Linear Koopman predictor.} With $\ph=\col(z,u)$, the IPE also attains the lowest median error on the academic example, $5.7\cdot10^{-3}$ against $6.0$ to $7.9\cdot10^{-3}$, and GeKo reduces it by $26$ to $31\%$ across seeds. On the robot, the linear predictor fails in all runs with Static and the RLS-type estimators, while IPE-data attains $3.0$\,cm without failures, which GeKo reduces by $48$ to $61\%$ across seeds. IPE-iso and IPE-data perform alike on all systems, so the invariant prior matters more than its shape.

\emph{Computational cost.} With the GeKo predictor, the estimator update takes on average $0.03$ to $0.11$\,ms per sample on all systems, due to the single $p\times p$ covariance. Solving \eqref{eq:mpc} takes on average $6$ to $9$\,ms for the robot and $4$ to $5$\,ms for the power system in MATLAB, which is well below the sampling periods of $50$\,ms and $20$\,ms, despite prediction horizons of $20$ and $30$, respectively.

\section{Conclusions}
We have developed an adaptive Koopman predictive controller that does not require persistent excitation by means of an invariant-prior estimator for matrix-valued operators. Using the offline GeKo operator as the mean of a matrix normal invariant prior yields a simple recursion with a single covariance matrix. For any regressor sequence, this covariance remains between two multiples of the prior covariance and the estimation error is input-to-state stable. On an academic example, a differential-drive robot and a power system, invariant-prior estimation attained the lowest median tracking error among the compared estimators, while its covariance remained in the band of Lemma~\ref{lem:cov} and the RLS covariance wound up. Prediction error bounds for robust predictive control, closed-loop stability and more advanced learning mechanisms are subjects of future research.
\acks{The author acknowledges the use of Claude (Anthropic) and ChatGPT (OpenAI) for literature searches and MATLAB code generation for the results in Section~\ref{sec:numerics}. The author reviewed and tested the code and maintains full responsibility for the code output.}
\bibliography{references}
\end{document}